\documentclass[runningheads]{llncs}
\usepackage{graphicx}
\usepackage{color,soul}
\usepackage{braket}
\usepackage{amsmath}
\usepackage{amsfonts}

\usepackage{subfig}

\begin{document}
%
\title{A Joint Framework to Privacy-Preserving Edge Intelligence in Vehicular Networks \thanks{This research was supported by the Republic of Korea’s MSIT (Ministry of Science and ICT), under the ICT Convergence Industry Innovation Technology Development Project(2022-0-00614) supervised by the IITP and partially supported by the Republic of Korea’s MSIT (Ministry of Science and ICT), under the 2022 technology commercialization capability enhancement project (2022-BS-RD-0034) supervised by the INNOPOLIS}}

\titlerunning{A Joint Framework to Privacy-Preserving Edge Intelligence in VNs}
%
\author{Muhammad Firdaus\inst{1}\orcidID{0000-0003-0104-848X} \and \\
Kyung-Hyune Rhee\inst{2}\orcidID{0000-0003-0466-8254}}
\authorrunning{M. Firdaus and K.H. Rhee}
%
\institute{Department of Artificial Intelligence Convergence, Pukyong National University, Busan 48513, Republic of Korea \\
\email{mfirdaus@pukyong.ac.kr}\\ \and
Division of Computer Engineering, Pukyong National University, Busan 48513, Republic of Korea \\
\email{khrhee@pknu.ac.kr}}

%
\maketitle              
%

\begin{abstract}
The number of internet-connected devices has been exponentially growing with the massive volume of heterogeneous data generated from various devices, resulting in a highly intertwined cyber-physical system. Currently, the Edge Intelligence System (EIS) concept that leverages the merits of edge computing and Artificial Intelligence (AI) is utilized to provide smart cloud services with powerful computational processing and reduce decision-making delays. Thus, EIS offers a possible solution to realizing future Intelligent Transportation Systems (ITS), especially in a vehicular network framework. However, since the central aggregator server is responsible for supervising the entire system orchestration, the existing EIS framework faces several challenges and is still potentially susceptible to numerous malicious attacks. Hence, to solve the issues mentioned earlier, this paper presents the notion of secure edge intelligence, merging the benefits of Federated Learning (FL), blockchain, and Local Differential Privacy (LDP). The blockchain-assisted FL approach is used to efficiently improve traffic prediction accuracy and enhance user privacy and security by recording transactions in immutable distributed ledger networks as well as providing a decentralized reward mechanism system. Furthermore, LDP is empowered to strengthen the confidentiality of data sharing transactions, especially in protecting the user’s private data from various attacks. The proposed framework has been implemented in two scenarios, i.e., blockchain-based FL to efficiently develop the decentralized traffic management for vehicular networks and LDP-based FL to produce the randomized privacy protection using the IBM Library for differential privacy.

\end{abstract}

\section{Introduction}
Recently, the number of internet-connected devices has been exponentially growing with great potential utilization in myriad applications, such as Intelligent Transportation Systems (ITS) \cite{ref29}, smart grids \cite{ref25}, smart healthcare \cite{ref22}, and smart industry \cite{ref2}. It is followed by the massive volume of heterogeneous data generated from various devices, resulting in a highly intertwined cyber-physical system. In terms of ITS, the concept of edge intelligence \cite{ref28} system (EIS), which leverages the merits of Mobile Edge Computing (MEC) and artificial intelligence (AI) technology, has been widely deployed to form the next generation of vehicular networks (VNs). MEC offers real-time communications with high bandwidth and low latency by locating the computing and processing infrastructure close to the end-user in the edge network. On the other hand, AI provides smart cloud services with high performance and reduces decision-making delays \cite{ref5}. Thus, EIS is designed to manage intelligent resource orchestration, enable self-aggregating communication systems, offer powerful computational processing, and reduce decision-making delay by leveraging edge resources on local edge networks \cite{ref30}.

Nevertheless, the traditional AI technique, such as machine learning, suffers from severe privacy leakage risk by centralizing and aggregating the user's training data containing private information on a centralized server. Federated learning (FL) as a decentralized machine learning paradigm has lately developed to address the privacy challenges by allowing mobile devices (e.g., vehicles) to collaboratively perform AI training without giving raw data containing the user's private information to the central aggregator \cite{ref11}. The FL approach allows the users to perform a local training model that never leaves their own devices. In this sense, the user’s raw data is only used to train and update a current global model and send an updated model to the central aggregator in each iteration. Then, the central aggregator generates a new global model by aggregating these updated and trained models gathered from the participated users to be used in the next iteration. This process is repeated in multiple iterations until the global model achieves a particular accuracy \cite{ref1}.

Although FL brings several advantages for edge intelligence systems, the existing FL framework still potentially experiences various adversarial attacks, such as membership inference and poisoning attacks \cite{ref20}. Here, in the membership inference attack, attackers might perform reverse engineering to gather user's private data by leveraging the updated model training, whereas a poisoning attack aims to affect the global model by sending the malicious updated models during the collaborative training phase. Furthermore, the central aggregator that is responsible for managing the whole system orchestration has trouble addressing crucial challenges associated with a Single Point of Failure (SPoF) issue, which may result in the whole FL system failure and lead to the risk of exposing private data. As a result, the users might be hesitant to participate in improving FL-based edge intelligence systems for VNs.

In order to address the above challenges, this paper presents the notion of secure edge intelligence, merging the benefits of Federated Learning (FL), blockchain, and Local Differential Privacy (LDP). The blockchain-assisted FL approach is used to enhance user privacy and security by recording transactions in immutable distributed ledger networks as well as improving an efficient traffic prediction accuracy in a decentralized manner. Moreover, blockchain can be deployed as an incentive mechanism to motivate the users to enhance the global model using their local data collaboratively. We also consider empowering the LDP technique to guarantee the confidentiality of the local training model by shielding the user’s data from malicious attackers. Thus, LDP supports heightening the edge intelligence’s data sharing transactions protection in VNs.

The structure of this paper is arranged as follows: we explain the background knowledge related to edge intelligence technology components in Section 2. Then, section 3 explains the related works. Next, Section 4 presents our proposed model, a secure edge intelligence on VNs. We discuss numerical results in Section 5. Finally, we conclude this paper in Section 6.

\section{Background} \label{sect:techoverview}
\subsection{Distributed Ledger Technology}
\label{subsec:blockchain}

\begin{figure}[t]
	\begin{centering}
	\includegraphics[width=0.6 \textwidth]{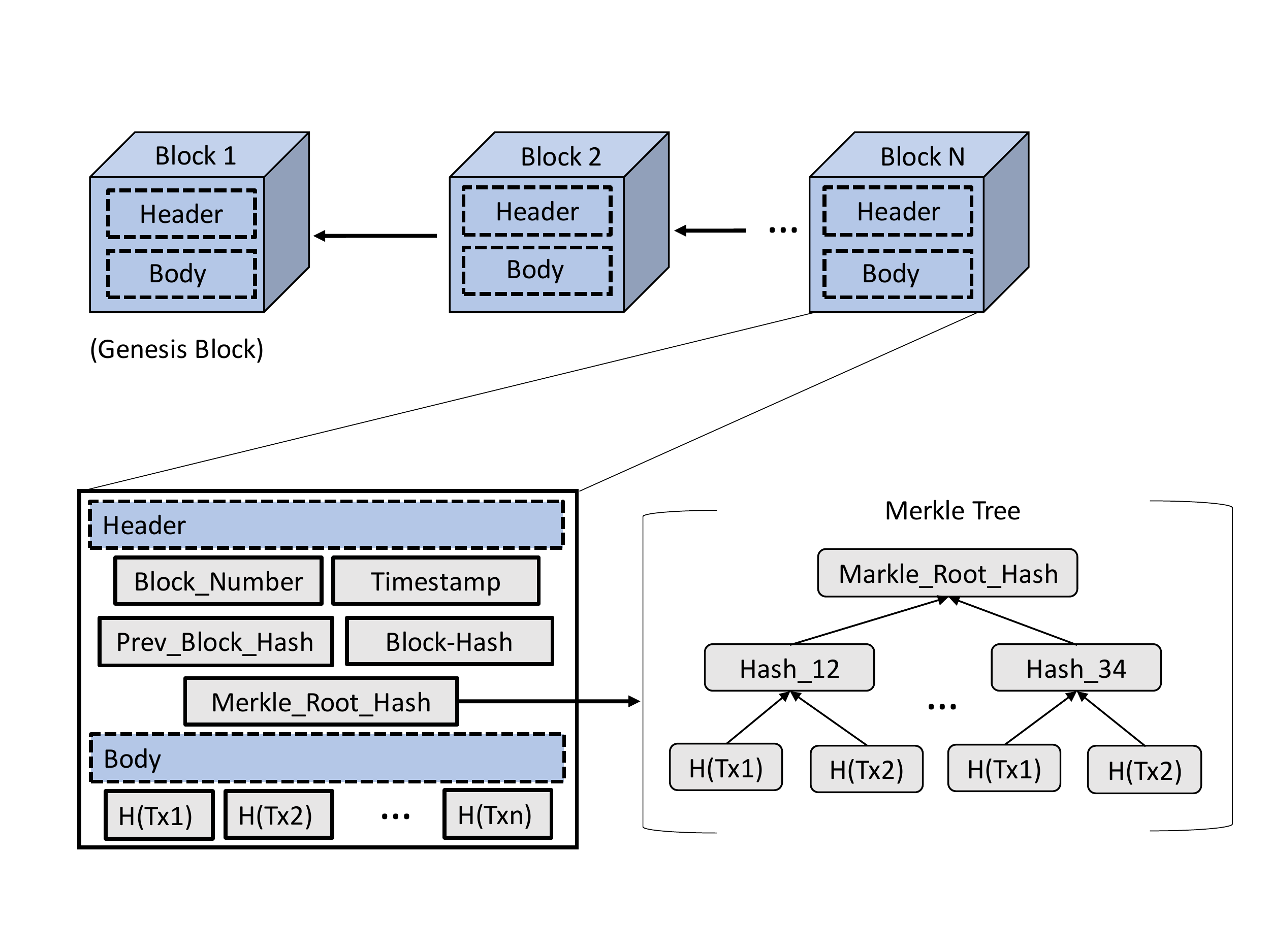}
	\caption{The illustration of blockchain structure}
	\label{fig:block}
	\end{centering}
\end{figure}

Since the introduction of Bitcoin in 2009 by Satoshi Nakamoto \cite{ref18}, blockchain has lately been earning attention from industry researchers and academia to develop decentralized and secure systems for various use cases. Blockchain can be utilized to address the bottleneck of a centralized server. It is an open database that supports anonymous and trustworthy transactions in ensuring data security without requiring any intermediaries. Here, all transactions are marked with a timestamp and recorded into a decentralized ledger where no single authority can endorse events secretly. Blockchain-enabled implementations generally take advantage of the SPoF feature with various consensuses mechanisms that validate and store verified transactions. Figure \ref{fig:block} shows the illustration of blockchain structure, Figure 1 shows the illustration of blockchain structure. The block is fundamentally composed of two parts, i.e., the block header, which contains information such as block number, previous hash block, Merkle tree root \cite{ref21}, timestamp, etc., and the block body includes the number of user's transactions in the blockchain network.

\subsection{Federated Learning} \label{subsec:FL}
The standard machine learning techniques utilize the training data to train the models by centralizing and aggregating the user's training data containing private information on a centralized server. Nevertheless, these approaches suffer from severe privacy risks, such as the potential of sensitive data leakage, the risk of SPoF, and enormous overhead in collecting and storing the training data. In order to address these issues, Google introduced federated learning as a promising method that permits distributed mobile devices to collaboratively train the models without centralizing the training data and keeps the local data stored on mobile devices.  As the user participants, each mobile device downloads the global model from a model provider (i.e., the central server), generates the model update by training the current global model using their local data, and then uploads them to the aggregator server. Then, as an aggregator server, the central server gathers and aggregates all the model updates from the user participants to produces a new global model for the next iteration. Thus, FL significantly enhances mobile device privacy by blocking some attacks for straightforward access to the local training data \cite{ref30}. Further, in the context of VNs, the federated learning method can be utilized to train prediction models without straightforward access to the private data on the vehicles, which protects the data privacy of vehicles and improves traffic prediction accuracy \cite{ref14}.

Basically, FL aims to facilitate the training model collaboration among participants without conveying their private data; thus, the private or confidential data is kept and never leaves their devices \cite{ref9}. The FL strives to optimize a global loss function $F(w)$ through an FL optimization objective that can be calculated using the empirical risk minimization approach in Equation 1,

\begin{align}
    \min_w{F(w)}=\sum_{k=1}^{m}p_k F_k(w)
    \label{eq:1}
\end{align}

where $w$, $m$, $p_k$, and $F_k(w)$, is notation for model parameters, number of devices, number of data points of device \textit{k} compared to total number of data points, and loss function of device \textit{k}, respectively.

\subsection{Differential Privacy}
\label{subsec:TEE}

Differential privacy (DP) \cite{ref6} has received much attention as a solution to the privacy-preserving challenges in machine learning \cite{ref19}. By including random noise, such as Gaussian or Laplacian noise distribution, DP offers a significant standard for data privacy protection. The participants manage the degree of privacy budget ($\epsilon$) that defines the number of noises added. Below is a description of the formal definition of DP \cite{ref6}.
\\
\textit{“A randomized mechanism M provides ($\epsilon,\delta$)- differential privacy if for any two neighboring database $D_1$ and $D_2$ that differ in only a single entry, $\forall S \subseteq Range (M)$,}

\begin{align}
    Pr(M(D_1) \in S) \leq e^\epsilon Pr(M(D_2) \in S)+\delta
    \label{eq:2}
\end{align}

if $\delta=0$, $M$ is said to achieve $\epsilon$-differential privacy.

In order to permit a slight possibility of failure, the term $\delta$ is denoted. The less value of $\epsilon$ (i.e., more additional noise) yields a better level of privacy, while the increased value of $\epsilon$ generates a lower level of privacy, according to equation \ref{eq:2}.

\section{Related Work}
\label{sect:relatedwork}

An edge computing (EC) network extends the notion of cloud computing concept to perform its capability to the network's edge. The main aims of EC are almost similar to cloudlets or fog computing in other references. Moreover, EC offers data storage, and computational processing is locally performed in the edge infrastructure to be closer to the data provider or user. As a result, edge computing offers real-time services, location-aware, and low-latency communication. Further, it also reduces delay and saves the bandwidth of transferring data for the remote node in the vehicular network system.
Moreover, current efforts utilized FL to improve the usability of MEC in reaching edge intelligence systems for wireless networks. In \cite{ref23}, the authors explored an FL model over wireless networks to improve FL's activities utilizing a control algorithm approach by respecting energy consumption as well as communication and computation latency. The authors in \cite{ref3} focus on improving system performance and solving the FL loss function problem during the training phase by optimizing resource allocation and user selection mechanisms. Further, the authors in \cite{ref17} deployed FL to form collaborative edge intelligence in mitigation situations of vehicular cyber-physical systems for detecting data leakage and protecting user's privacy information.

Blockchain as a distributed ledger technology is suggested to address the weakness of a traditional data management system in VNs. In the FL-based EIS, blockchain can be used to provide a decentralized incentive mechanism, verify the trustworthiness of the updated model training, and support a fair global model's aggregation. Lately, some research has offered to merge FL and blockchain to strengthen privacy. In \cite{ref16}, the authors suggested a privacy-preserving data sharing mechanism in industrial IoT for a distributed multi-parties scenario. They integrate the consensus mechanism of permissioned blockchain with FL. Meanwhile,  the work from \cite{ref15} designed the framework to refuse dishonest users from the FL system by automatically enforcing smart contracts to defy model or data poisoning attacks. Furthermore, the authors in \cite{ref27} proposed the DeepChain protocol that employs a blockchain-based incentive mechanism to provide a secure, auditable, fair, and distributed deep learning system. Here, the incentive is utilized to force participants to act rightly and substitute a centralized approach's drawbacks.

In order to guarantee the confidentiality of the local training model from malicious attackers, some study concentrates on employing differential privacy for users' data privacy protection. In \cite{ref24}, the authors proposed a hybrid approach to tackle the problem of inference attacks and provide privacy-preserving federated learning using differential privacy and secure multi-party computation (SMPC). This approach aims to address the FL challenges, such as inference attacks and lack of accuracy, as well as recede the enlargement of noise injection when the number of users rises in diverse use cases and applications. Further, the study in \cite{ref26} proposed the NbAFL framework to evade data leakage using a differential privacy technique by adding noise before FL model aggregation. This study focuses on solving the information leakage in the distributed stochastic gradient descent (SGD) based FL and develops a theoretical convergence bound for the trained FL model's loss function. 

\section{Towards Secure Edge Intelligence}
\label{sect:proposed-model}

\begin{figure}[t!]
	\begin{centering}
	\includegraphics[width=0.8\textwidth]{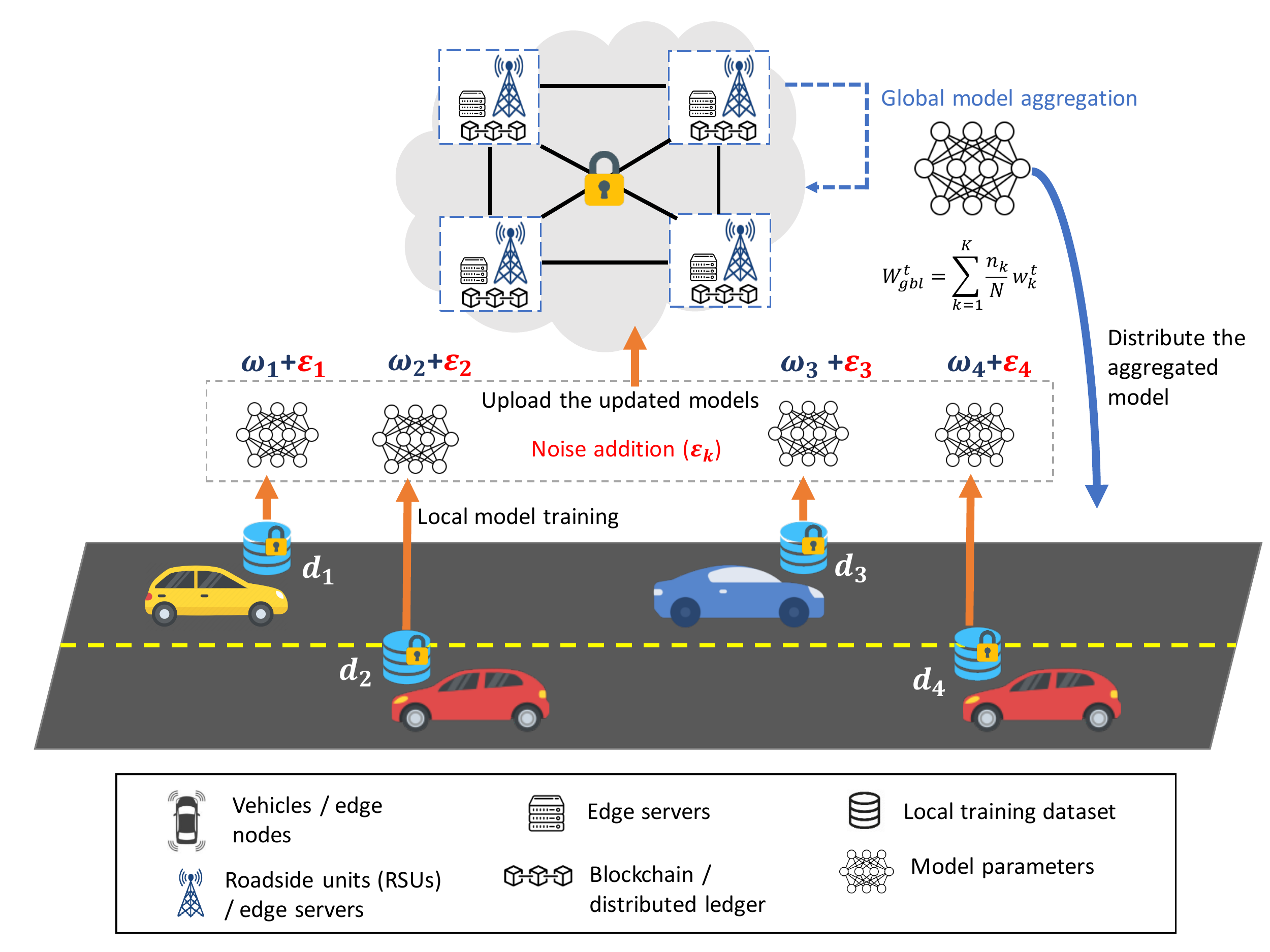}
	\caption{Design architecture of secure edge intelligence on IoVEC.}
	\label{fig:architecture}
	\end{centering}
\end{figure}

As illustrated in Figure \ref{fig:architecture}, we propose the joint framework by leveraging the advantages of FL, LDP, and blockchain technology to form a secure edge intelligence in VNs. We use the blockchain to enhance the privacy and security of model parameters in the edge resource of federated learning by encrypting the data with a particular cryptography technique. Moreover, blockchain as a distributed ledger technology effectively overcame a centralized server's drawbacks and handled the uploaded parameters of updated models transparently. Furthermore,  LDP is empowered to strengthen the confidentiality of transactions, primarily in defending the sensitive or private user's data on the trained local model uploading process.

In this architecture, vehicles and roadside units (RSUs) are the primary nodes, act as the user's participants and aggregator server, respectively. They communicate with each other by forming vehicle-to-vehicle (V2V) and vehicle-to-infrastructure (V2I) communication. These two types of communications refer to dedicated short-range communication (DSRC) standards that facilitate single or multi-hop communication among VNs entities \cite{ref12}. Here, we consider vehicles as distributed edge users and utilize their local data to train FL models. Also, they are equipped with simple communication and computation capabilities supported by onboard units (OBUs), consisting of various sensing devices. On the other hand, RSUs are equipped with edge computing servers and designed as the distributed edge servers stationed along the road, providing wireless communications from roadside infrastructure to vehicles. Further, RSUs are considered as intelligent edge servers, providing and aggregating global models from distributed edge users in VNs.

Our proposed design architecture comprises three parts: the local data training executed by vehicles, model parameter validation and protection by empowering LDP-based blockchain, and global aggregation in the distributed edge aggregator server. First, the system is started with the initial learning model process, where initial model parameters of the global model $W^0$ are uploaded to the blockchain-empowered distributed RSUs. Then, the edge users (i.e., vehicles, donated by $k$) in iteration $t$ retrieve global parameters $W^t$ from blockchain and execute local training to generate the updated models $w_k^t$ using their local dataset $d_k$ based on equation \ref{eq:1}. Then, the LDP mechanism is conducted by adding random noise $\epsilon$ to the updated models $w_k^t$ to strengthen the privacy during uploading the trained local model and defend against linkability attacks, such as membership inference attacks. In this case, $k$ adds noise $\epsilon_k$ to 	achive $\epsilon$-differential privacy based on equation \ref{eq:2} using Gaussian mechanism, which is defined by: 

\begin{align}
    f(D)+N(0, S_f^2\sigma^2)
    \label{eq:3}
\end{align} where  $N(0,S_f^2 \sigma^2)$ is the normal distribution with mean 0, and standard deviation $S_f\sigma$ \cite{ref7}. 

After that, $k$ uploads $w_k^t$ with $\epsilon$ to the blockchain over distributed RSUs. In short, through this phase, vehicles train their dataset locally and upload the trained model updates collaboratively. Later, a particular consensus mechanism verifies and aggregates $w_k^t$ to obtain a new global model $W_{gbl}^t$ for the next iteration ($t+1$), where:

\begin{align}
    W_{gbl}^t=\sum_{k=1}^{K}\frac{n_k}{N} w_k^t
    \label{eq:4}
\end{align} where $n_k$ is the number of samples generated by $k$ and $N$ is the total number of data points (samples). Thus, the iteration continues until the model reaches a precise accuracy or the number of iterations exceeding the upper limit. Therefore, in this model, the edge servers (i.e., RSUs) maintain the blockchain and legitimate for performing the global aggregation process to generate a new global model in the VNs.

\begin{figure}[t!]
	\begin{centering}
	\includegraphics[width=0.7\textwidth]{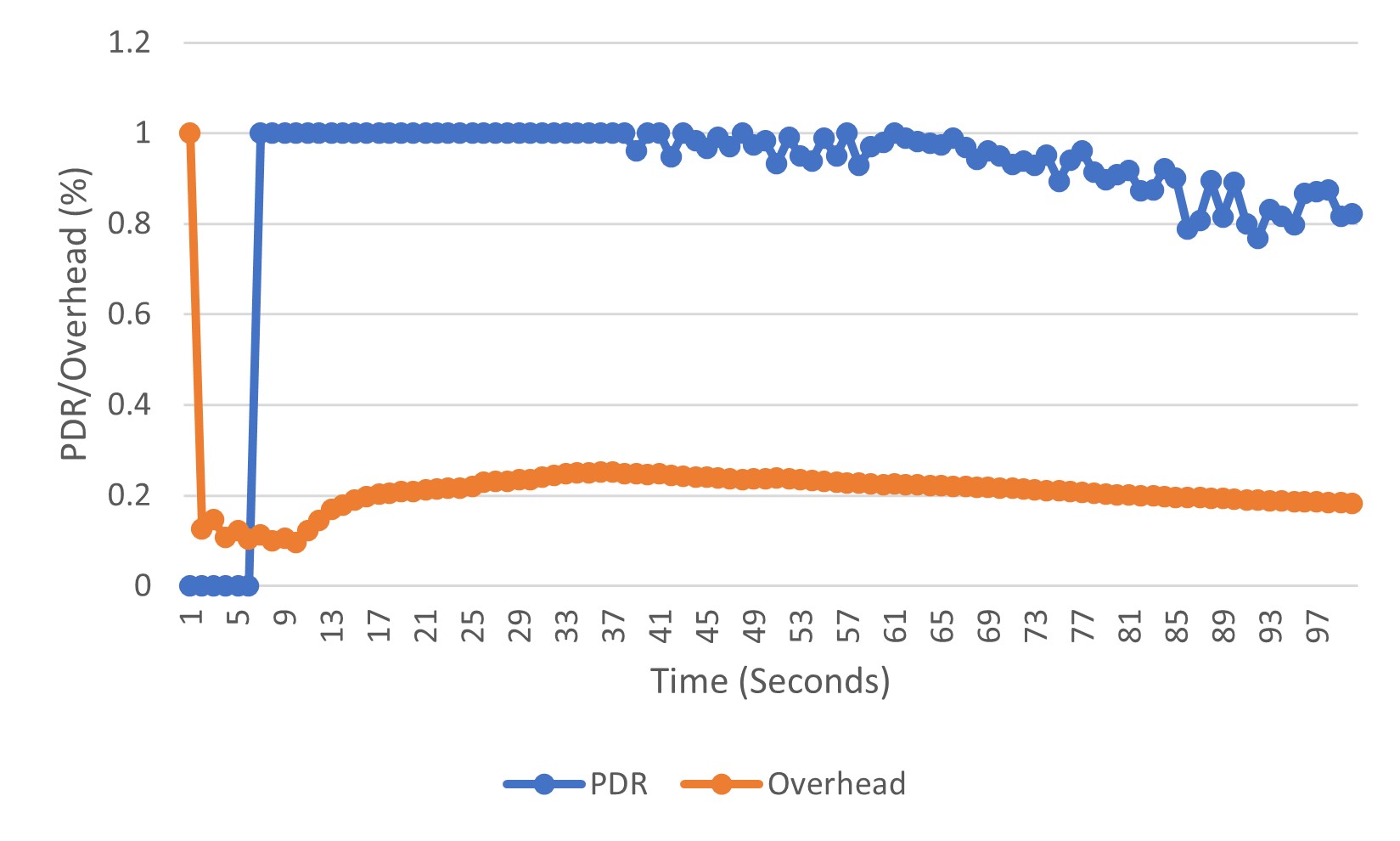}
	\caption{Packet delivery ratio against MAC/PHY overhead.}
	\label{fig:pdr/overhead}
	\end{centering}
\end{figure}

\begin{figure}[t!]
	\begin{centering}
	\includegraphics[width=0.55\textwidth]{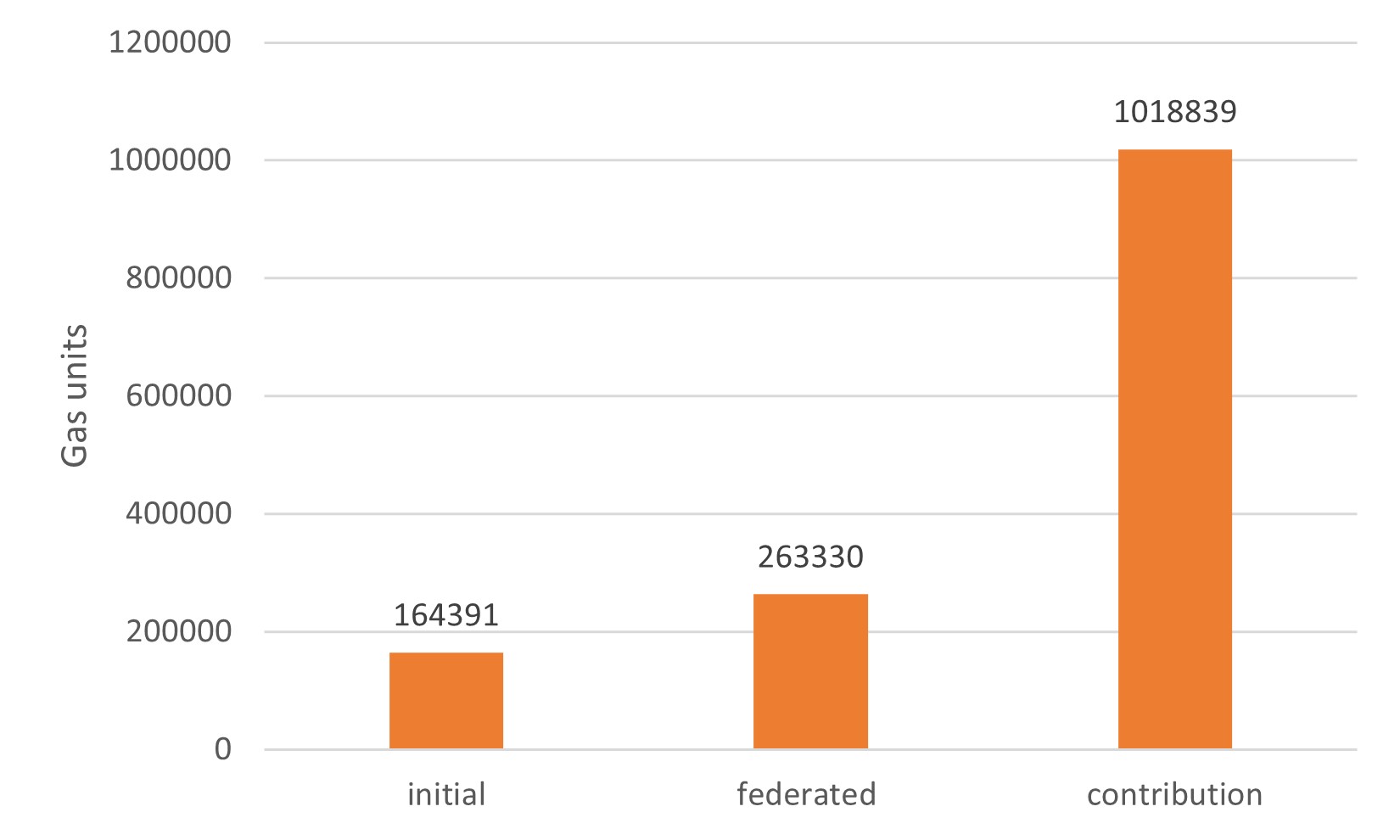}
	\caption{Initial migration and deploying smart contracts.}
	\label{fig:migration}
	\end{centering}
\end{figure}

\begin{figure}[t!]
\centering
\subfloat[$\epsilon$=4.03 and $\delta$=1e-05]{{\includegraphics[width=.6\textwidth]{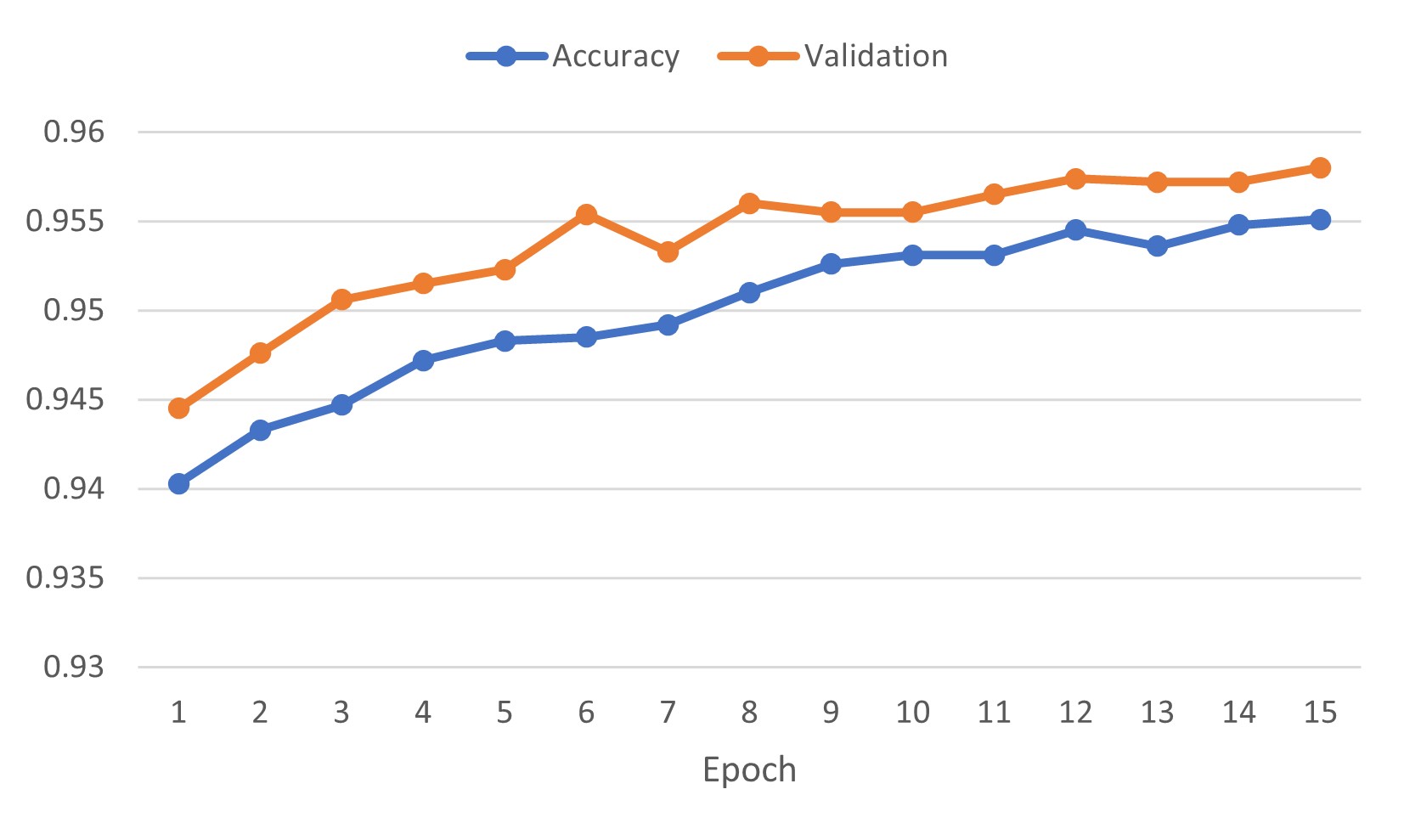} }\label{fig:mycaption-a}}

\subfloat[$\epsilon$=1.18 and $\delta$=1e-05]{{\includegraphics[width=.6\textwidth]{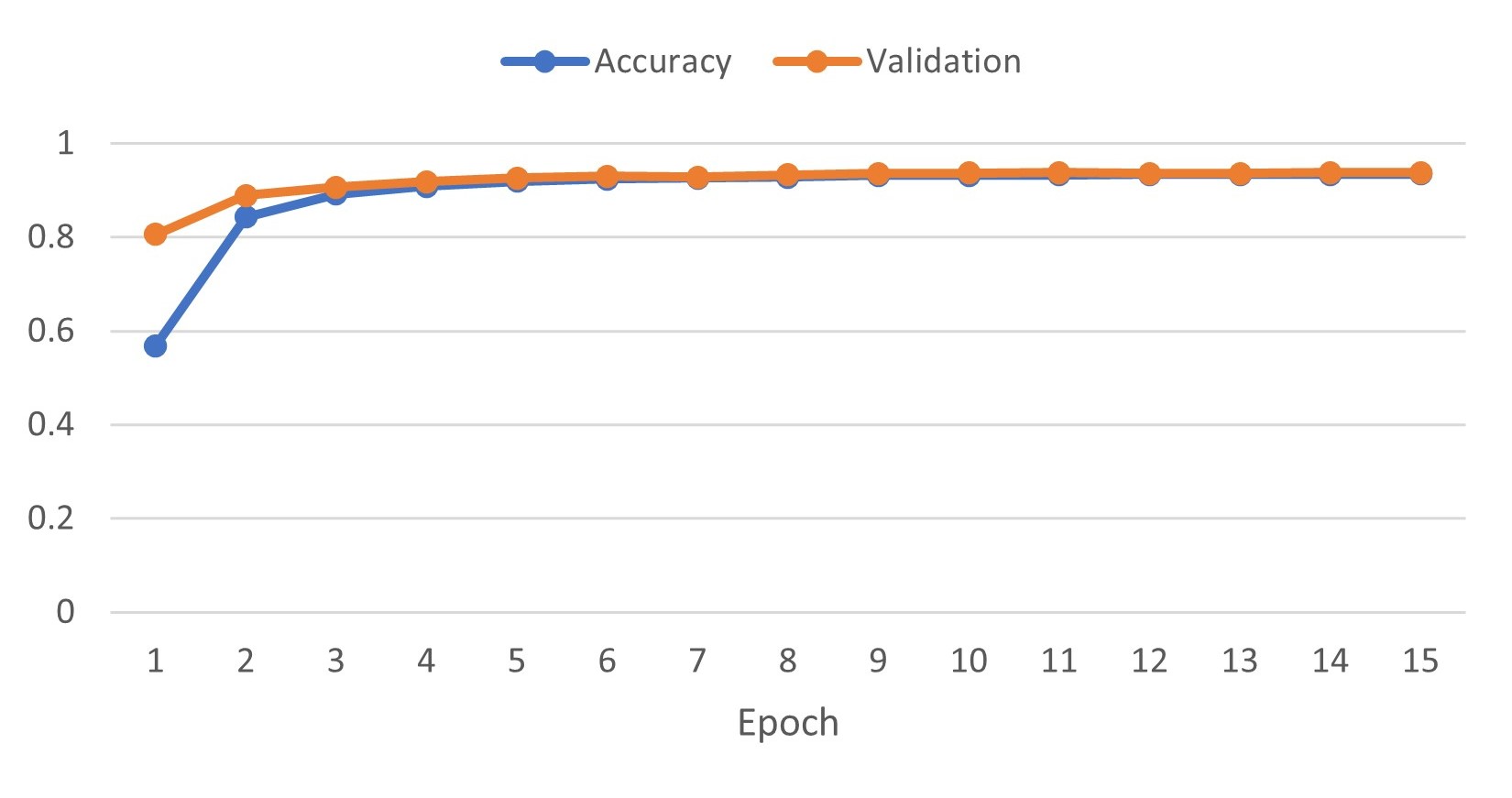} }\label{fig:mycaption-b}}

\subfloat[$\epsilon$=0.522 and $\delta$=1e-05]{{\includegraphics[width=.6\textwidth]{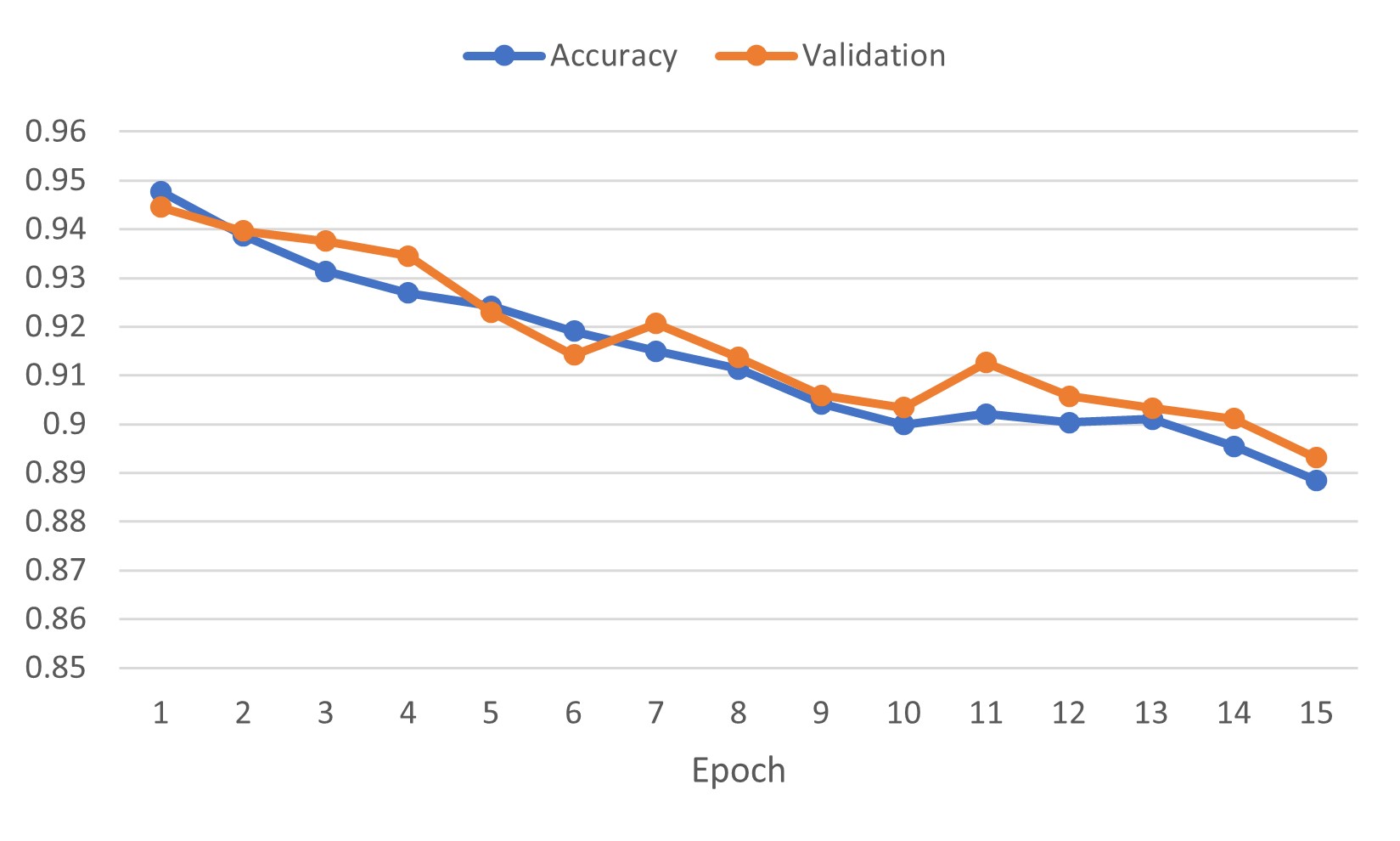} }\label{fig:mycaption-c}}

\caption{Privacy budget $\epsilon$ against accuracy}
\label{fit:mycaption}
    \end{figure}

\section{Numerical Results and Discussion}\label{sect:discussion} 

The proposed framework has been implemented in two scenarios, i.e., blockchain-based FL to efficiently develop the decentralized traffic management for vehicular networks and LDP-based FL to produce the randomized privacy protection using the IBM Library for differential privacy. The vehicular network prototype was designed using a discrete event simulator with an optimized link-state routing protocol to analyze network performance; the detailed prototype can be seen in our previous work \cite{ref8}. Figure \ref{fig:pdr/overhead} shows the Packet Delivery Ratio (PDR) against MAC/PHY overhead during a simulation time of 100 seconds. As seen in the figure, after 17s of simulation time, the overhead is practically consistent within the range of 0.2 to 0.25, and it even gradually decreases. The lower the vehicle overhead, the greater the performance of the system, and vice versa. Our proposed protocol is relatively efficient based on the preceding results because it does not incur a significant amount of overhead.

In order to form a decentralized FL system based on blockchain technology, we use the consortium setting \cite{ref3} that leverages blockchain to realize the decentralized FL transaction, transparently evaluate the participants' contributions to the global model, and develop a decentralized incentive system. In this experiment, MNIST \cite{ref13} datasets are used as a standard image classification with 10.000 images for the testing examples and 60.000 images for the training examples. Each example is a 28 × 28 size gray-level image. Figure \ref{fig:migration} shows the initial migration and deploying smart contracts using the Ethereum platform. The figure shows that we need 164391 (0.00328782 ETH), 263330 (0.0052666 ETH), and 1018839 (0.02037678 ETH) units of gas for initial migration, federated smart contract, and deploying participants' contribution implementation, respectively. After deploying smart contracts, we can customize the number of participants involved for local model training and calculate their contribution fairly based on blockchain. Then, to implement the FL with DP model, we use a python-based open-source library that IBM developed for the experimentation, simulation, and deployment of differential privacy tools and applications \cite{ref10}. Figure \ref{fit:mycaption} shows DP-based FL experiment using different privacy budget degrees, i.e., $\epsilon=4.03$, $\epsilon=1.18$ and $\epsilon=0.522$, in 15 epochs. According to simulation results, Figure \ref{fig:mycaption-a} with privacy budget $\epsilon=4.03$ achieves 95.80000024915344 model accuracy. In contrast, Figure \ref{fig:mycaption-b} and  \ref{fig:mycaption-c} with privacy budget $\epsilon=1.18$ and $\epsilon=0.522$, generates a model accuracy of 93.77999901771545 and 89.31000232696533, respectively. Thus, the smaller value of $\epsilon$ (i.e., more noise added) generates higher privacy (i.e., based on the gap between accuracy and validation) but lower accuracy, and vice versa.

\section{Conclusion and Future Work}
In this paper, we presented the notion of secure edge intelligence for the VNs by leveraging the merits of FL, blockchain, and LDP.  We use the blockchain to overcome a centralized server's drawbacks. Moreover, blockchain can be used to form a decentralized incentive system to encourage participants to share their trained model. Furthermore, we use LDP to strengthen the confidentiality of data sharing transactions, especially in protecting the user's private data from various attacks. However, even though the FL approach is a promising method to implement in a decentralized system, there are some significant challenges, especially in user selection issues for the model training process as well as system and statistical heterogeneity. Further, we need to consider the effect of the privacy budget $\epsilon$ on the accuracy in future work, where the smaller value of $\epsilon$ (i.e., more noise added) generates higher privacy but lower accuracy. Thus, it is essential to consider these challenges for future research direction.
%
%
%

\begin{thebibliography}{}
\bibitem[1]{ref1}
Tran The Anh et al. “Efficient training management for mobile crowd-machine learning: A deep reinforcement learning approach”. In: IEEE Wireless Communications Letters 8.5 (2019), pp. 1345–1348.

\bibitem[2]{ref2}Hugh Boyes et al. “The industrial internet of things (IIoT): An analysis framework”. In: Computers in industry 101 (2018), pp. 1–12.

\bibitem[3]{ref3}Harry Cai, Daniel Rueckert, and Jonathan Passerat-Palmbach. “2cp: Decentralized protocols to transparently evaluate contributivity in blockchain federated learning environments”. In: arXiv preprint arXiv:2011.07516 (2020).

\bibitem[4]{ref4}Mingzhe Chen et al. “A joint learning and communications framework for federated learning over wireless networks”. In: IEEE Transactions on Wireless Communications 20.1 (2020), pp. 269–283.

\bibitem[5]{ref5}Yueyue Dai et al. “Artificial intelligence empowered edge computing and caching for internet of vehicles”. In: IEEE Wireless Communications 26.3 (2019), pp. 12–18.

\bibitem[6]{ref6}Cynthia Dwork. “Differential privacy: A survey of results”. In: International conference on theory and applications of models of computation. Springer. 2008, pp. 1–19.

\bibitem[7]{ref7}Cynthia Dwork, Aaron Roth, et al. “The algorithmic foundations of differential privacy.” In: Found. Trends Theor. Comput. Sci. 9.3-4 (2014), pp. 211–407.

\bibitem[8]{ref8}Muhammad Firdaus and Kyung-Hyune Rhee. “On blockchain-enhanced secure data storage and sharing in vehicular edge computing networks”. In: Applied Sciences 11.1 (2021), p. 414.

\bibitem[9]{ref9}Andrew Hard et al. “Federated learning for mobile keyboard prediction”. In: arXiv preprint arXiv:1811.03604 (2018).

\bibitem[10]{ref10}Naoise Holohan et al. “Diffprivlib: the IBM differential privacy library”. In: arXiv preprint arXiv:1907.02444 (2019).

\bibitem[11]{ref11}Jiawen Kang et al. “Reliable federated learning for mobile networks”. In: IEEE Wireless Communications 27.2 (2020), pp. 72–80.

\bibitem[12]{ref12}John B Kenney. “Dedicated short-range communications (DSRC) standards in the United States”. In: Proceedings of the IEEE 99.7 (2011), pp. 1162–1182.

\bibitem[13]{ref13}Yann LeCun et al. “Gradient-based learning applied to document recognition”. In: Proceedings of the IEEE 86.11 (1998), pp. 2278–2324.

\bibitem[14]{ref14}Le Liang, Hao Ye, and Geoffrey Ye Li. “Toward intelligent vehicular networks: A machine learning framework”. In: IEEE Internet of Things Journal 6.1 (2018), pp. 124–135.

\bibitem[15]{ref15}Yi Liu et al. “A secure federated learning framework for 5G networks”. In: IEEE Wireless Communications 27.4 (2020), pp. 24–31.

\bibitem[16]{ref16}Yunlong Lu et al. “Blockchain and federated learning for privacy-preserved data sharing in industrial IoT”. In: IEEE Transactions on Industrial Informatics 16.6 (2019), pp. 4177–4186.

\bibitem[17]{ref17}Yunlong Lu et al. “Federated learning for data privacy preservation in vehicular cyber-physical systems”. In: IEEE Network 34.3 (2020), pp. 50–56

\bibitem[18]{ref18}Satoshi Nakamoto. “Bitcoin: A peer-to-peer electronic cash system”. In: Decentralized Business Review (2008), p. 21260.

\bibitem[19]{ref19}Reza Shokri and Vitaly Shmatikov. “Privacy-preserving deep learning”. In: Proceedings of the 22nd ACM SIGSAC conference on computer and communications security. 2015, pp. 1310–1321.

\bibitem[20]{ref20}Reza Shokri et al. “Membership inference attacks against machine learning models”. In: 2017 IEEE symposium on security and privacy (SP). IEEE. 2017, pp. 3–18.

\bibitem[21]{ref21}Michael Szydlo. “Merkle tree traversal in log space and time”. In: International Conference on the Theory and Applications of Cryptographic Techniques. Springer. 2004, pp. 541–554.

\bibitem[22]{ref22}Shuo Tian et al. “Smart healthcare: making medical care more intelligent”. In: Global Health Journal 3.3 (2019), pp. 62–65.

\bibitem[23]{ref23}Nguyen H Tran et al. “Federated learning over wireless networks: Optimization model design and analysis”. In: IEEE INFOCOM 2019-IEEE Conference on Computer Communications. IEEE. 2019, pp. 1387–1395.

\bibitem[24]{ref24}Stacey Truex et al. “A hybrid approach to privacy-preserving federated learning”. In: Proceedings of the 12th ACM workshop on artificial intelligence and security. 2019, pp. 1–11.

\bibitem[25]{ref25}Fadi Al-Turjman and Mohammad Abujubbeh. “IoT-enabled smart grid via SM: An overview”. In: Future Generation Computer Systems 96 (2019), pp. 579–590.

\bibitem[26]{ref26}Kang Wei et al. “Federated learning with differential privacy: Algorithms and performance analysis”. In: IEEE Transactions on Information Forensics and Security 15 (2020), pp. 3454–3469.

\bibitem[27]{ref27}Jiasi Weng et al. “Deepchain: Auditable and privacy-preserving deep learning with blockchain-based incentive”. In: IEEE Transactions on Dependable and Secure Computing 18.5 (2019), pp. 2438–2455.

\bibitem[28]{ref28}Zhi Zhou et al. “Edge intelligence: Paving the last mile of artificial intelligence with edge computing”. In: Proceedings of the IEEE 107.8 (2019), pp. 1738–1762.

\bibitem[29]{ref29}Li Zhu et al. “Big data analytics in intelligent transportation systems: A survey”. In: IEEE Transactions on Intelligent Transportation Systems 20.1 (2018), pp. 383–398.

\bibitem[30]{ref30}Xudong Zhu, Hui Li, and Yang Yu. “Blockchain-based privacy preserving deep learning”. In: International Conference on Information Security and Cryptology. Springer. 2018, pp. 370–383.

\end{thebibliography}

%


\end{document}